\documentclass[pre,aps,twocolumn,superscriptaddress,nofootinbib]{revtex4-1}
\usepackage{graphicx}
\usepackage{amssymb,latexsym,amsthm,amsmath}    
\usepackage{mathtools}
\usepackage{float}
\usepackage[final]{changes}
\usepackage{hyperref}
\usepackage{color}
\usepackage{colortbl}

\begin{document}
\title{High-order couplings in geometric complex networks of neurons}

\author{A. Tlaie}
\affiliation{Complex Systems Group \& GISC, Universidad Rey Juan Carlos, 28933 M\'ostoles, Madrid, Spain}
\affiliation{Center for Biomedical Technology, Universidad Polit\'ecnica de Madrid, 28223 Pozuelo de Alarc\'on, Madrid, Spain}
\affiliation{Department of Applied Mathematics and Statistics, ETSIT Aeron\'auticos, Universidad Polit\'ecnica de Madrid, 28040 Madrid, Spain }
\author{I.  Leyva}
\affiliation{Complex Systems Group \& GISC, Universidad Rey Juan Carlos, 28933 M\'ostoles, Madrid, Spain}
\affiliation{Center for Biomedical Technology, Universidad Polit\'ecnica de Madrid, 28223 Pozuelo de Alarc\'on, Madrid, Spain}
\author{I.  Sendi\~na-Nadal}
\affiliation{Complex Systems Group \& GISC, Universidad Rey Juan Carlos, 28933 M\'ostoles, Madrid, Spain}
\affiliation{Center for Biomedical Technology, Universidad Polit\'ecnica de Madrid, 28223 Pozuelo de Alarc\'on, Madrid, Spain}

\begin{abstract}
We explore the consequences of introducing higher-order interactions in a geometric complex network of Morris-Lecar neurons. We focus on the regime where travelling synchronization waves are observed out of a first-neighbours based coupling, to evaluate the changes induced when higher-order dynamical interactions are included. We observe that the travelling wave phenomenon gets enhanced by these interactions, allowing the information to travel further in the system without generating pathological full synchronization states. This scheme could be a step towards a simple modelization of neuroglial networks.


\end{abstract}
\maketitle
\date{}

\section{Introduction}

The combination of complex networks  and non-linear dynamics has provided a solid framework for the study of a large number of very different real systems that can be analysed as large ensembles of dynamical units with non trivial connectivity patterns; these systems are as diverse as economics \cite{souma2003complex}, genetics \cite{sieberts2007moving},  social dynamics \cite{fellman2014modeling}, or neuroscience \cite{bullmore2009complex}.

Among all the possible collective features that can emerge in this context, synchronization is the most extensively studied, since it has been revealed as the fundamental mechanism in the transmission of  information 
in all kinds of dynamical ensembles \cite{Arenas2008}. One of the fields where this perspective has led to new research lines is in neuro scientific applications. The neural system can be considered as a dynamical complex network in all its relevant scales, ranging  from the microscopic, where the networked elements are single neurons \cite{eckmann2007physics,kumar2010spiking}, through the cortical column mesoscale \cite{zeng2018mesoscale}, to the entire brain \cite{eguiluz2005,bullmore2009complex}, with the  brain areas acting as nodes of a functional network defined in terms of correlation levels.

 However, even if synchronization is a key mechanism involved in the  coordination of  the neural ensemble, it is well known that exceedingly high levels of synchronization can destroy the overall complexity of the system, reducing its ability to process  information  and, eventually, leading to pathological states as epilepsy \cite{jiruska2013synchronization}. Therefore, a healthy synchronous functioning  in the brain needs the existence of mechanisms of regulation, both structural and dynamical, to ensure the proper equilibrium between coordination and function segregation. 

A plausible regulating mechanism is the glial ensemble, whose role in the brain performance  is a long-standing problem in neuroscience. At the microscale, it is known that astrocytes can establish up to $10^5$ synapses, meaning that they might be responsible for the modulation of the electrical response of neurons sharing  no anatomical connection at all \cite{Astrocitos1, Astrocitos2, Astrocitos3,Pitta2016astrocytes} and, therefore, they could be the source of   high-order interactions supporting coordination levels that overcome the outreach of direct neural connectivity.  

Several attempts have been made to model the neuroglial interaction \cite{depitta2019computational},  most of them focusing on the neuron-astrocyte pair or, more commonly, a triade of two neurons and an astrocyte \cite{sajedinia2018new, li2016astrocytic}. However,  few studies have considered the  networked context,  which is usually mathematically and computationally costly  \cite{amiri2012functional}.  In this work we model the neural-glial ensemble as a geometrical network with synaptic coupling, where the synaptic modulation of astrocytes is introduced using a high-order interaction formalism developed by Estrada et al. in Refs. \cite{Estrada1,Estrada2,Estrada3,Estrada4}. It provides a solid quantitative mean to simulate and analyze the dynamics of a system in which these higher-order interactions are present. These effects are susceptible of revealing themselves particularly important in space embedded systems, where the Euclidean distances shape not only the probability of connection but also their weights. The high-order connectivity operator allows us to extend the usual first-neighbour interaction scheme, that disregards higher-order interactions under the implicit assumption that if two nodes are not topologically connected they do not dynamically interact; such an assumption  is no longer valid in a network of neurons whose communication is mediated by astrocytes. 

The application of the high-order connectivity formalism to a complex network of synaptically connected neurons can provide insights about how introducing not only first but also second-neighbors interactions might be useful to comprehend further details of the neuronal dynamics in a simple and mathematically well-defined way. We show how it enhances the appearance of synchronization waves, a mean for transmitting information throughout the system in a coherent way, but avoiding the neuronal hyper-synchronization disorder that would result from increasing a direct neural connectivity. 

\section{Model}\label{sec:ML}

The network consists of an ensemble of $N$ neurons that are randomly seeded in a $2D$ Euclidean square area of size $L \times L$. The nodes are connected following a distance-dependent geometric rule, such that neuron $i$ has a probability of establishing a link with neuron $j$ \cite{Kaiser2004, barthelemy2011spatial, Leyva2011}:  
\begin{equation}\label{linkprob}
p_{ij} = p_0 \; e^{-\left(\frac{r_{ij}}{l_c}\right)}
\end{equation}
where $p_0$ is a normalization constant, $r_{ij}$ is the Euclidean distance between $i$ and $j$, and $l_c$ the correlation length parameter that controls the typical outreach of the connections when constructing the network; low values of this parameter yield highly clustered, short-ranged networks, while  standard Erd\"os-R\'enyi networks are obtained in the limit of large $l_c$. The neural connectivity is encoded in the correspondent adjacency matrix ${ A}=\{ a_{ij} \}$ such that $a_{ij}=1$ means a physical connection between neurons $i$ and $j$ and $a_{ij}=0$ otherwise.

Single node dynamics is implemented as a Morris-Lecar (ML) neuron \cite{morris1981}:
\begin{eqnarray}
C \dot{V_i} &=&  -g_{\rm Ca} M_{\infty}( V_i-V_{\rm Ca} ) - g_{\rm K} W_i( V_i-V_{\rm K} ) 
\\ &-& g_{\rm l} ( V_i-V_{\rm l}) + I_i + I^{\rm ext}_i , \nonumber \\
\dot{W_i} &=& \phi \,\tau_W  ( W_{\infty}-W_i ) \nonumber
\label{ML}
\end{eqnarray}

\noindent where $V_i$ and $W_i$ are, respectively, the membrane potential and the fraction of open $\rm K^+$ channels of the $i$th neuron ; $\phi$ is a reference frequency. The parameters $g_{\rm X}$ and $V_{\rm X}$ account for the electric conductance and equilibrium potentials of the $\rm X=\{K,Ca,\text{leaky}\}$ channels. An external current $I_i^{\rm ext}=I_0+Q\xi_i$ is added, with $I_0=50$ mA  chosen such that neurons are sub-threshold to their natural firing regime, which, in this case, will be induced by the additive white Gaussian noise $Q\xi_i$ of zero mean and intensity $Q$.
 
Additionally, the channel voltage-dependent saturation values $M_{\infty}, W_{\infty}$, $\tau_W$ respond to hyperbolic functions dependent on $V_i$:
\begin{eqnarray}
M_{\infty}(V_i) &=& \frac{1}{2}\left[1+\tanh\left(\frac{V_i-V_1}{V_2}\right)\right] \label{eq:M} \\
W_{\infty}(V_i) &=& \frac{1}{2}\left[1+\tanh\left(\frac{V_i-V_3}{V_4}\right)\right] \label{eq:W}\\
\tau_W(V_i) &=& \cosh \left( \frac{V_i-V_3}{2V_4} \right) \label{eq:tau}
\end{eqnarray}
The explicit value of every parameter can be found in Table~ \ref{tab:Parameters}.

 \begin{table}[h]
\centering
\begin{tabular}{ |c|c| }
 \hline
 $C$ & $20.0$ $\mu \text{F/cm}^2$ \\ 
 \hline
 $g_{Ca}$ & $ 4.0$ $\mu \text{S/cm}^2 $ \\ 
 \hline 
 $g_K$ & $ 8.0$ $\mu \text{S/cm}^2 $ \\ 
 \hline
 $g_l$ & $ 2.0$ $\mu \text{S/cm}^2 $ \\   
 \hline 
 $V_{Ca}$ & $ 120.0 $ mV  \\  
 \hline
 $V_K$ & $-80.0$ mV  \\   
 \hline
 $V_l$ & $ -60.0$ mV  \\   
 \hline
 $V_1$ & $ -1.2$ mV  \\  
 \hline 
 $V_2$ & $ 18.0$ mV  \\  
 \hline 
 $V_3$ & $2.0$ mV\\  
 \hline 
 $V_4$ & $ 17.4$ mV  \\  
 \hline 
 $\phi$ & 1/15 \\
 \hline
$Q$ & 0.5 mA \\
 \hline
\end{tabular}
\centering
\caption{Parameters used for the Morris-Lecar simulations.}
\label{tab:Parameters}
\end{table}

The direct synaptic interaction between presynaptic $j$ neuron and excitatory postsynaptic $i$ neuron  is captured by the injected current $I_{ij}$ \cite{Leone2013,Mofakham2016,Esfahani2016}:
\begin{equation}\label{eq:Ii2}
 I_{ij}= \frac{\sigma}{K} \left[  e^{-2(t-t_j)} (V_0-V_i) \right],
\end{equation}
 with $t_j$ being the time of the last spike of node $j$. The synaptic conductance $\sigma$, normalized by the largest node degree $K$ (number of connections that a given node has) present in the network, plays the role of coupling strength. In the local coupling approximation,  first-order neighbours contribute to the synaptic coupling, and therefore $I_{i}=\sum_{j\in {\cal N}_i} I_{ij}$, where ${\cal N}_i$ is the neighbourhood of node $i$, that is, nodes $j$ such that $a_{ij}=1$. 
 
 In spatial, highly clustered networks with reduced link range $l_c$, the coupling configuration described in Eq.~(\ref{eq:Ii2}) favors travelling wave synchronization, as long as $\sigma$ is high enough \cite{Leyva2011}. On the contrary, in the mean field approximation limit, $l_c \to L$, only globally incoherent/coherent states are accessible \cite{Rosenbaum2014}. In our model, as a balance between these two extrema, we intend to explore the potential enhancement effect of higher-order connectivity at the local spatial scale, as for example the glial ensemble has in the neural circuits that are not directly connected among them \cite{Pitta2016astrocytes}. Therefore, following the same mathematical framework developed in Ref.~\cite{Estrada1,Estrada2, Estrada3}, we allow that the injection current $I_i$  accounts for the contribution not only from neurons $j\in {\cal N}_i$ whose topological distance  is $d_{ij}=1$, but also from neighbors at higher topological distances $d_{ij}>1$., that is: 
 \begin{equation}\label{eq:Ii}
 I_i= \sum_{d=1}^{D} d^{-\alpha} \left( \sum_{j | d_{ij}=d} I_ {ij} \right),
\end{equation}
where $D$ is the maximal topological distance considered. The successively distant contributions to the injection synaptic current $I_i$ are modulated by a geometrically decaying term, $d^{-\alpha}$, where the constant  $\alpha$ is a suppression parameter for the distance-dependent coupling strength. Notice that when the summation is limited to the first order $D=1$, the usual first-neighbours interaction is recovered.

\section{Synchronization measures}\label{sec:Sync}

In order to quantify the level of coordination among the network firing events we count how many neurons fire within the same time window. First, the total simulation time $T$ is divided in $N_b$ bins of a convenient size $\tau$, longer than the time duration of each individual spike, but much shorter than the average inter-spike interval (ISI). Then, the total simulation time is discretized as $T=N_b\tau$ and the time series of the dynamics of  neuron $i$th is replaced by the binary series $B_i$, where $B_i(n)=1$ if the $i$th neuron spiked within the $n$th time bin, and $0$ otherwise, with $n=1,\dots,N_b$. This simplification of the dynamics ensures a fast an precise calculation of the ensemble statistics. Finally, the coherence between the spiking sequence of neurons $i$ and $j$  can be characterized with the quantity $s_{ij}\in [0,1]$
\begin{equation}
s_{ij}=\frac{\sum_{n=1}^{N_b}B_i(n)B_j(n)}{\sum_{n=1}^{N_b}B_i(n)\sum_{n=1}^{N_b}B_j(n)},
\end{equation}
\noindent where the term in the denominator is a normalization factor, such that $s_{ij}=1$ implies full coincidence between the spike trains of neurons $i$ and $j$. The ensemble average of $s_{ij}$ is the global synchronization measure $S$, given by:
\begin{equation}
S=\langle s_{ij} \rangle = \frac{1}{N(N-1)}\sum_{i\neq j}^N s_{ij} 
\end{equation} 

However, as we are interested not only in global but also spatial local effects in the ensemble coherence from the high-order couplings, we use the coherence matrix $s_{ij}$ to compute also the {Euclidean} local synchronization $S_\rho$, where only the correlation values $s_{ij}$ of those neurons pairs which are closer to each other than a given distance $\rho$ are taken into account, that is \cite{Leyva2011}:
\begin{equation}
S_\rho = \langle s_{ij} \rangle, \; \forall \lbrace i,j \rbrace ~|~ r_{ij} < \rho
\end{equation}

It is expected that in the limit $\rho \to L$,  $S_{\rho} \sim S$. In the following, all measures are averaged over five realizations of different networks. 

\begin{figure}[t]
\includegraphics[width=\columnwidth]{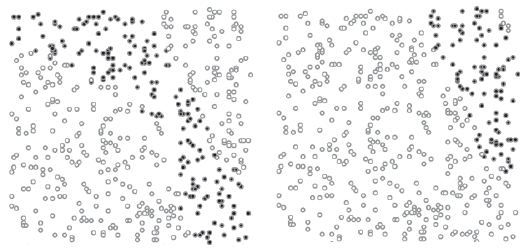}
\caption{Successive snapshots of the neurons' spiking activity in the travelling wave synchronization mode. Filled dots represent spiking neurons while empty dots represent silent neurons. In the example the wave is propagating from the left to the right in a network of $N=150$ Morris-Lecar neurons. Other parameters: $p_0=1.0$, $L=50$, $l_c=0.15$, $\sigma=150$, $D=1$. }
\label{fig_D2}
\end{figure}

\section{Results}

\begin{figure}
\includegraphics[width=\columnwidth]{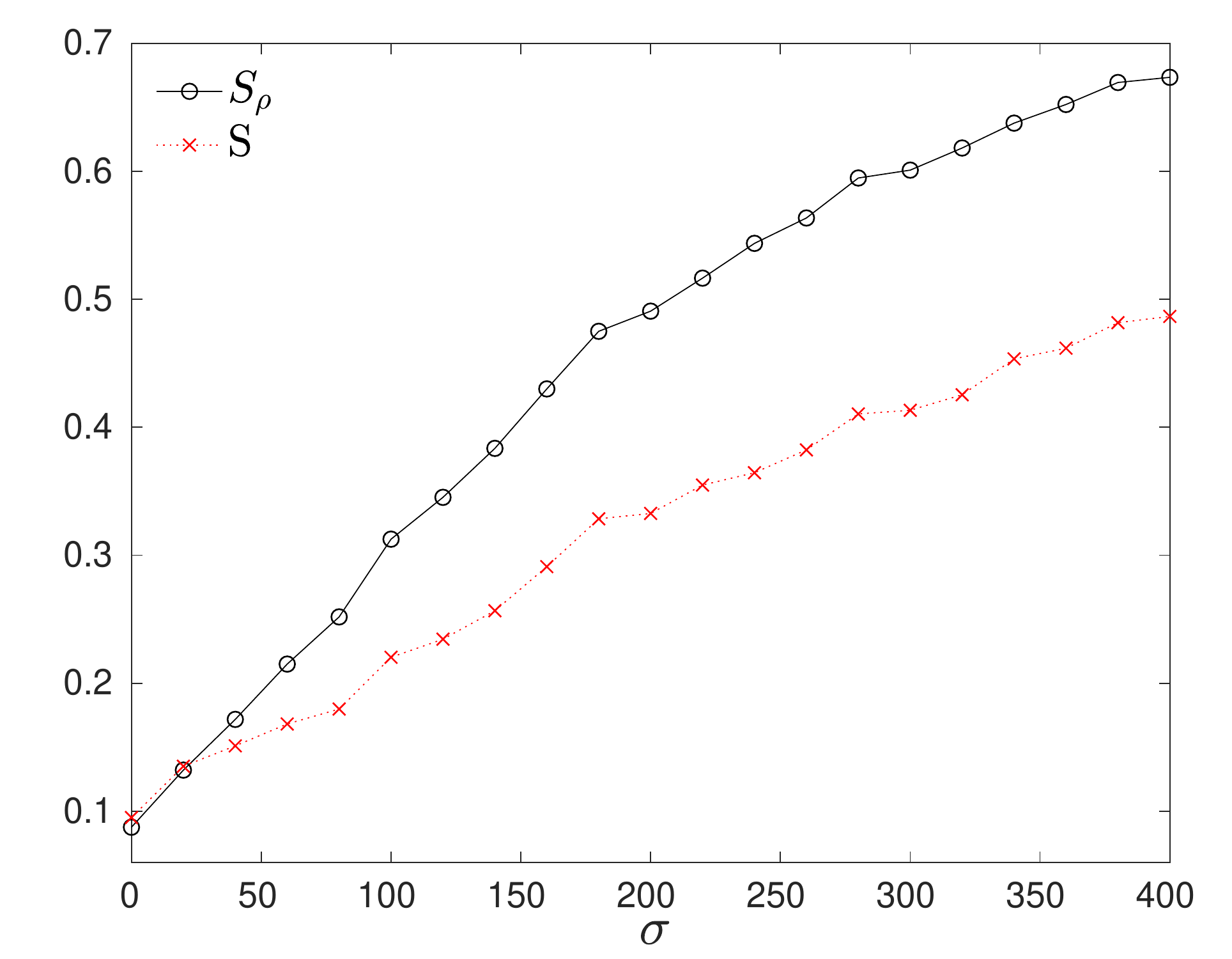}
\caption{(Color online) Synchronization route as a function of the coupling strength $\sigma$ for $l_c=0.15$ and $D=1$. Local synchronization $S_{\rho}$ (black circles) computed with $\rho=10$ and global synchronization (red crosses). Both series of data are averages over five $N = 150$ network realizations, with the connectivity scheme outlined in Sec.~\ref{sec:ML}.}
\label{fig_D1}
\end{figure} 

When just first-order interactions are present, this geometrical arrangement of neurons favours the propagation of travelling waves of neurons' spiking activity, supported by  a highly clustered structure with a typically low link outreach \cite{Leyva2011}. To illustrate such propagation, in Fig.~\ref{fig_D2} we show  two successive snapshots of an example where  $l_c=0.15$ and $\sigma=150$. Here black dots represent spiking neurons while void dots portray those which are silent. The links between nodes are not included for clarity. This feature is queantified in Fig.~\ref{fig_D1}, showing that this wave-like phenomenon is characterized by a local synchronization $S_{\rho}$ (circles) larger than the global synchronization $S$ (crosses), as it can be observed  The low value of the link outreach $l_c$ prevents the system to reach full synchronization even when the coupling strength $\sigma$ increases, whereas the local synchronization $S_\rho$ grows much faster, indicating a reinforcement of the wave activity.  

We now evaluate the effect of introducing higher order contributions in the synaptic coupling in Eq.~(\ref{eq:Ii}) received from neighbors at topological distance up to $D=2$. We compute the difference $S_{\rho}-S$ as a measure to quantify the existence of either a travelling wave front (when $S_{\rho}-S$ is large) or global or null synchronization ($S_{\rho}\sim S$) otherwise.

\begin{figure}
\includegraphics[width=\columnwidth]{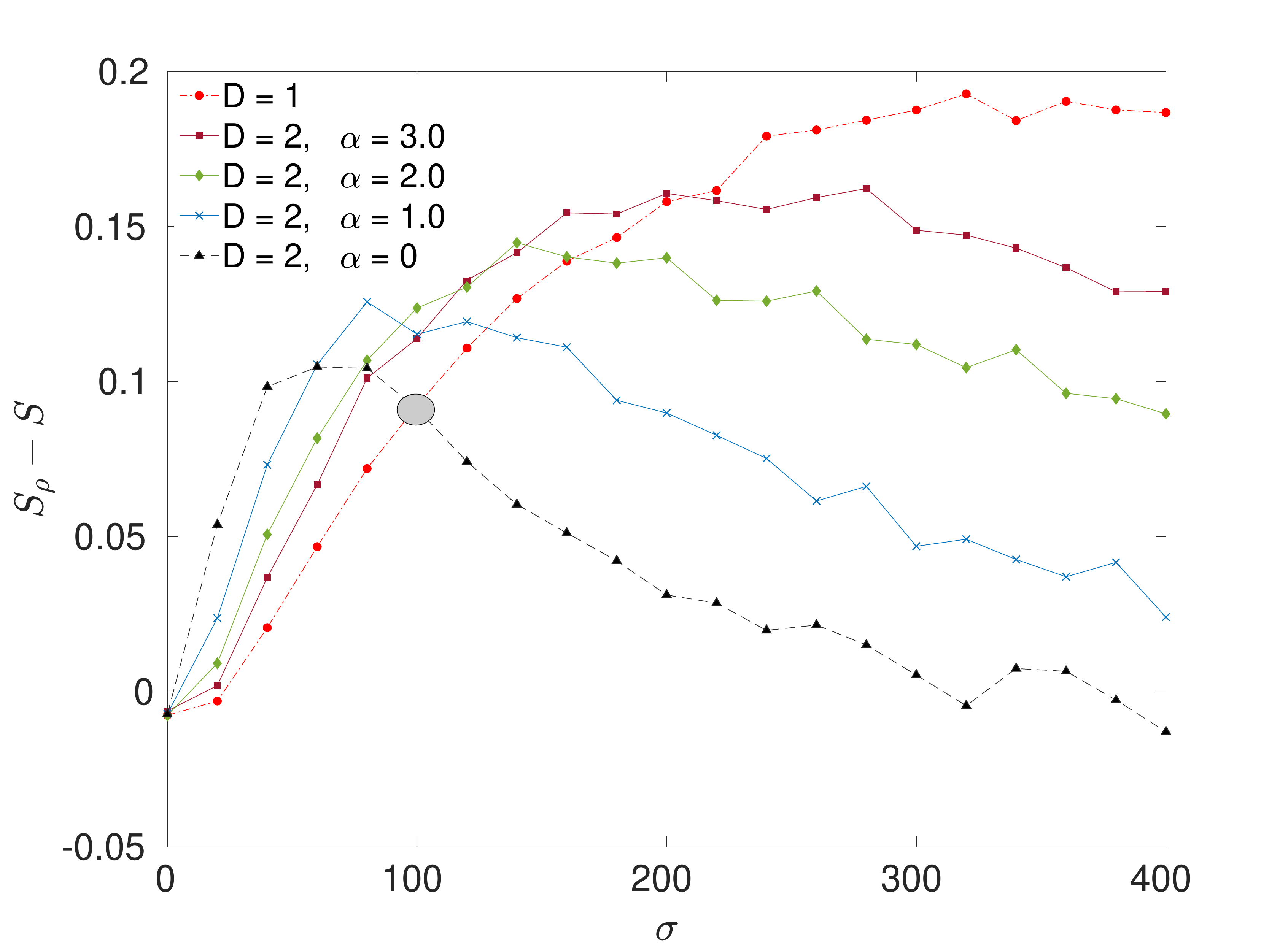}
\caption{(Color online)  Difference between local $S_{\rho}$ and global $S$ synchronization as a function of the coupling strength $\sigma$ for different values of the suppression constant $\alpha$ and maximal topological distances $D=1$ and $D=2$. The gray circle highlights the intersection between the curves $D=1$ and $(D=2,\alpha=0)$. Each point is an average over 5 network realizations. Same parameters as in Fig.~\ref{fig_D1}.}
\label{fig_D3}
\end{figure}

Results are collected in Fig.~\ref{fig_D3}, where $S_{\rho}-S$ is plotted as a function of the conductance $\sigma$ for different values of the suppression constant $\alpha$, ranging between $0$ and $3$. For the sake of comparison, the curve for $D=1$ is included (red circles). When $D>1$, the higher the value of the suppression $\alpha$, the weaker the influence from D = 2 neighbors. Therefore, we observe that for the higher suppression $\alpha = 3$ (purple squares), the behavior approaches the $D = 1$, and both curves overlap in almost the whole range of explored couplings up to $\sigma \sim 200$. 

However, as high-order effects become stronger for smaller values of $\alpha$ in Fig.~\ref{fig_D3}, the $S_\rho$-$S$ curves exhibit a maximum, located at lower values of the conductance $\sigma$. For instance, the curve for $D=2$, $\alpha=0$ (black triangles) peaks at $\sigma \sim 50$, while for the case $D=1$ the $S_{\rho}-S$ difference is very small. The conclusion is that the introduction of another layer of interacting neighbors allows the propagation of travelling waves for coupling strengths  where first-order interactions only supports incoherent activity.
It can be deduced that this critical value of $\sigma$ is related to the best communication efficiency of the spiking activity, given the constraints of a fixed topology and dynamical parameters. Notice that the travelling wave feature implies a temporal ordering of the network's activity, as opposed to global synchronization (no temporal order) or incoherent activity (random spikes) and, therefore, this dynamical regime ensures a robust encoding of information. 

\begin{figure}
\includegraphics[width=\columnwidth]{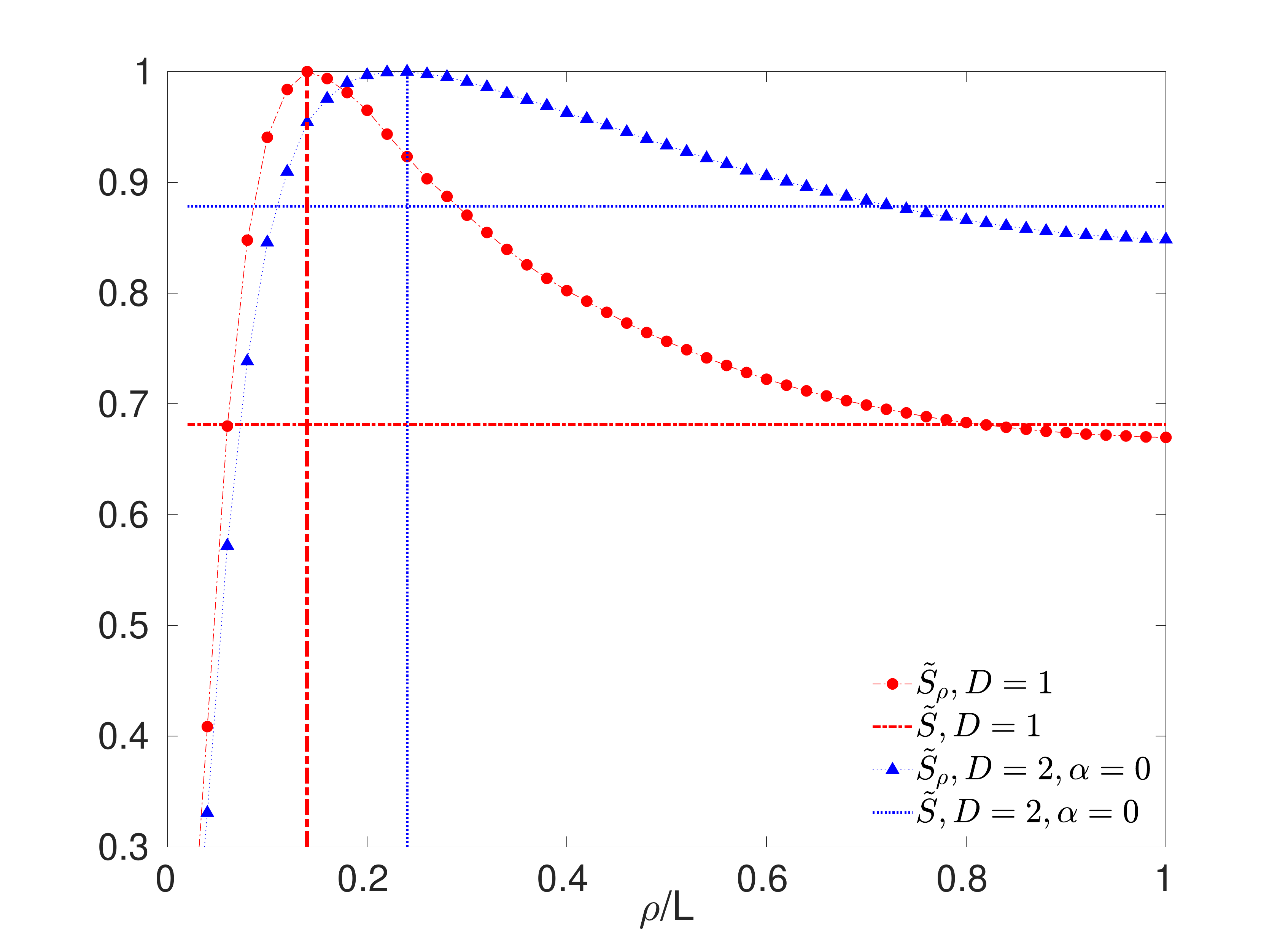}
\caption{(Color online)  Normalized local synchronization ($\tilde{S_{\rho}}$) values as a function of $\rho/L$ for the two sets of parameter conditions defined by the gray circle in Fig.~\ref{fig_D3}. Horizontal dashed lines mark the normalized values of the corresponding global synchronization. Vertical dashed lines mark the point defining the width of the travelling front. Each point is averaged over 10 network realizations.}
\label{fig4}
\end{figure}

To further explore the network activity and travelling waves features, we focused on the coupling strength at which the two previously mentioned curves intersect (grey circle in Fig. \ref{fig_D3}, $\sigma \sim 100$), corresponding in both cases to wave propagation. However, we can observe that the wavefront features are also modified by the high-order effects. We analyzed these differences by varying the scale $\rho$ at which the local synchronization $S_{\rho}$ is measured for both cases at the crossing point. Figure~\ref{fig4} compares $S_{\rho}$ for $D=1$ (red circles) and $D=2$, $\alpha$=0 (blue triangles),  normalized to their respective maxima, $\tilde{S_{\rho}}=S_{\rho}/\max(S_{\rho})$, as a function of $\rho/L$, such that, when $\rho$ is of the same size $L$ as the surface in which the network is seeded, the local synchronization statistically converges to the normalized global synchronization level ($\tilde{S} = S/max(S)$) observed for each case (horizontal dashed lines). As expected, there is an optimal length scale $\rho$ at which the local synchronization measure is maximum: smaller scales undervalue the cluster of neurons spiking synchronously, while larger scales average neurons which are in different dynamical states. Therefore, the value of $\rho$ at which $\tilde{S}_{\rho}$ peaks is an estimation of the wave-front width. Thus, as Fig.~\ref{fig4} indicates, higher-order interactions, for the same conductance value, allow the propagation of wider spiking waves, almost doubling the size with respect to $D=1$. This could lead us to conclude that taking into account the direct influence of neighbors at larger topological distance allows the information to be transmitted faster throughout the network, as more neurons are active in each wave front (while preserving the locality feature) and thus, the wave front needs less time to cross the entire network. 


\section{Conclusions}

In this work we have evidenced that the introduction of higher-order dynamical interactions in a ensemble of neurons with geometrical connectivity patterns leads to a faster and much more robust propagation of the information through this spatially embedded system. The propagation occurs as a travelling wave, whose wave-front gets enhanced thanks to recruiting more neurons in the transmission. In addition, we have shown that  higher-order dynamical interactions allows this kind of time-ordered synchronization for much lower coupling values than the case where only first-order neighbours are involved. 

We hypothesize that this could be an innovative way of modelling the neuro-glial interaction, among other physical systems in which higher-order interactions need to be taken into account. Specifically, we argue that this mechanism of higher-order interactions could be a potent and computationally cheaper approach to the alternatives that can be found nowadays in the literature. The central foundation for having chosen this particular mathematical formalism comes from a biological insight: astrocytes have been evidenced to modulate up to $\approx 10^5$ synapses \cite{Astrocitos4}, while the majority of the neurons they interact with do not share an anatomical connection. This would imply that, while there is a given number of topological links in the network, some indirect ones would be present in the form of dynamical modulation, this role being played by astrocytes. As this is only a first step towards modeling the interplay between astrocytes and neurons in a network, we focused on establishing a solid base upon which we will continue the research. 

Financial support from the Ministerio de Econom\'ia y Competitividad of Spain under project FIS2017-84151-P and from the Group of Research of Excelence URJC-Banco de Santander is acknowledged. A.T. acknowledges support from the Comunidad de Madrid through the European Youth Employment Initiative and the Rey Juan Carlos University. Authors acknowledge the computational resources and assistance provided by CRESCO, the supercomputing center of ENEA in Portici, Italy.

\end{document}